\documentclass[journal]{IEEEtran}
\usepackage{psfrag}
\usepackage{amsmath}
\usepackage{amssymb}
\usepackage{color}
\usepackage{graphicx}
\usepackage{epsfig}
\usepackage{fancybox}
\usepackage{subfigure}
\usepackage{cite}
\usepackage{algorithm}
\usepackage{algorithmic}
\usepackage{multirow}
\usepackage{theorem}
\usepackage{url}
\usepackage{epstopdf}
\usepackage{algorithm}
\usepackage{algorithmic}
\usepackage{color}
\usepackage{siunitx}
\usepackage{multicol}
\usepackage{mathtools}
\usepackage{epstopdf} 
\usepackage{setspace}
\usepackage{array}
\usepackage{siunitx}
\usepackage{soul}
\usepackage[hidelinks]{hyperref}

\graphicspath{ {Figs/} }
\DeclareMathSizes{9}{9}{9}{9}
\newcommand{\ignore}[1]{}

\newcommand\numberthis{\addtocounter{equation}{1}\tag{\theequation}}
\newcolumntype{L}[1]{>{\raggedright\let\newline\\\arraybackslash\hspace{0pt}}m{#1}}
\newcolumntype{C}[1]{>{\centering\let\newline\\\arraybackslash\hspace{0pt}}m{#1}}
\newcolumntype{R}[1]{>{\raggedleft\let\newline\\\arraybackslash\hspace{0pt}}m{#1}}

\begin{document}
\title{SkyLogic -- A proposal for a skyrmion logic device}
\author{Meghna G. Mankalale,~\IEEEmembership{Student Member, IEEE}, 
Zhengyang Zhao, \\ Jian-Ping Wang,~\IEEEmembership{Fellow, IEEE} and 
Sachin S. Sapatnekar,~\IEEEmembership{Fellow, IEEE}
 \thanks{All the authors are with the Department of Electrical and Computer Engineering, University of
 Minnesota, Minneapolis, MN, USA (email: \{manka018, zhaox526, jpwang,
sachin\}@umn.edu) }}
\maketitle

\begin{abstract}
This work proposes a novel logic device (SkyLogic) based on skyrmions, which are
magnetic vortex-like structures that have low depinning current density and
are robust to defects. A charge current
sent through a polarizer ferromagnet (P--FM) nucleates a skyrmion 
at the input end of an intra-gate FM interconnect with 
perpendicular magnetic anisotropy (PMA--FM). The output end
of the PMA--FM forms the free layer of an MTJ stack. 
A spin Hall metal (SHM) is placed beneath the PMA--FM. The skyrmion is propagated 
to the output end of the PMA--FM by passing a charge current through the SHM. 
The resistance of the MTJ stack is low (high) when
a skyrmion is present (absent) in the free layer, thereby realizing
an inverter. A framework is developed to analyze the performance 
of the SkyLogic device.  A circuit-level technique is developed that counters 
the transverse displacement of skyrmion in the PMA--FM and allows use 
of high current densities for fast propagation. The design space exploration of the PMA--FM 
material parameters is performed to obtain an optimal design point.  
At the optimal point, we obtain an inverter delay of $434$ps with a switching energy of $7.1$fJ.
\end{abstract}


\section{Introduction}
\label{sec:intro}

Recently, research in spintronics has accelerated in
an effort to find an alternative to or complement the existing
CMOS-based electronics. 
Several physical phenomena have been exploited to develop 
novel spin logic devices~\cite{pan2017expanded, BeyondCMOS2015}. 
Some of the more successful logic device concepts are based on
manipulation of the magnetic nanostructures like 
domain-walls~\cite{allwood2005magnetic,STMG,CoMET}, but 
domain-walls are susceptible to pinning due to material defects~\cite{pizzini2009pinning}. 
Recently, skyrmions, which are vortex-like 
spin structures in magnetic thin films, have been actively
studied~\cite{fert2013skyrmions}. Skyrmions have proven to be more robust to 
pinning compared to domain walls~\cite{fert2013skyrmions}. 
The recent room temperature experimental observation of skyrmion creation, 
current-driven displacement, and detection~\cite{fert2017review},
make them attractive structures to develop skyrmion-based
logic devices. 

\begin{figure}[ht]
\centering
\subfigure[]{
	\centering
	\includegraphics[width=8.25cm]{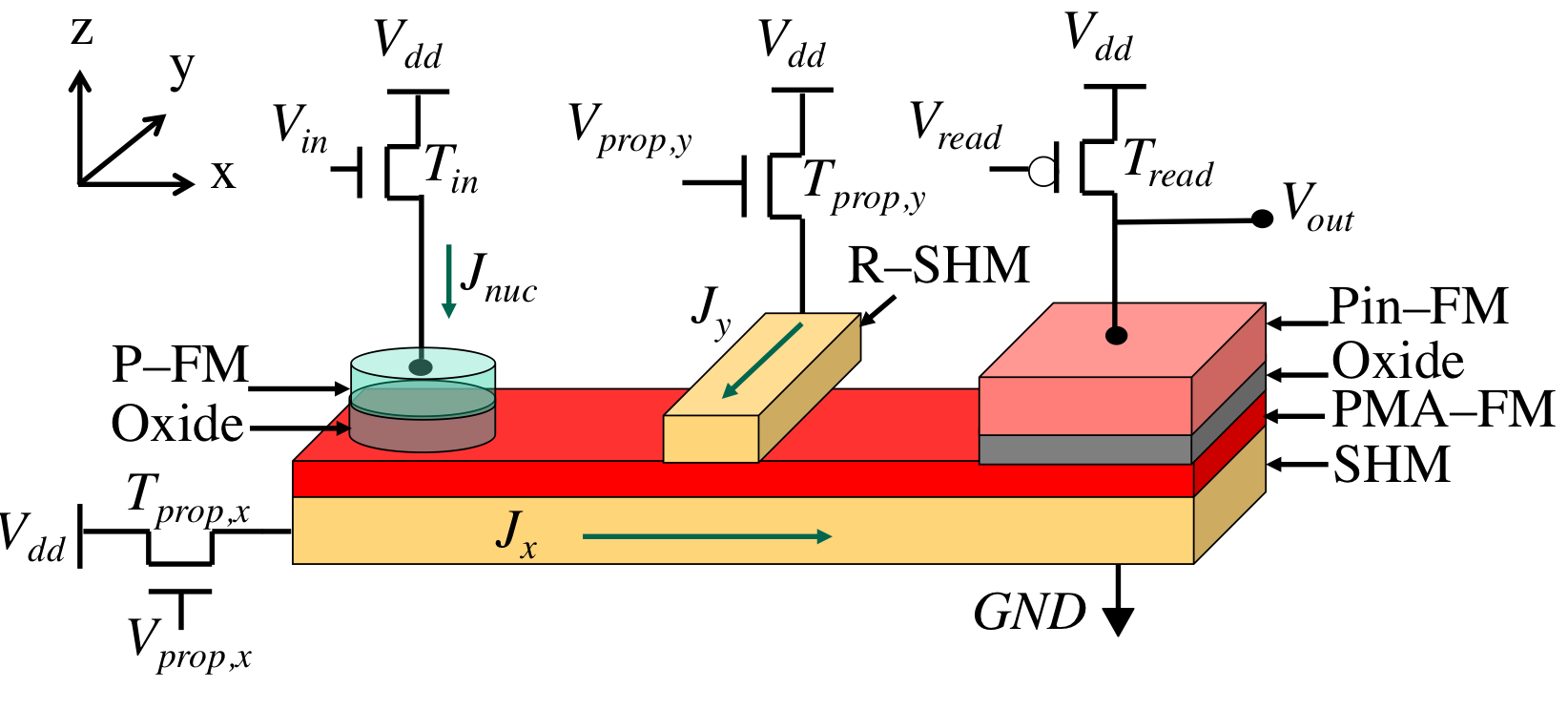}
}
\caption{(a) The structure of the SkyLogic device.}
\label{fig:device}
\end{figure}

Several skyrmion-based logic devices~\cite{weishangSkyrmionSpectrum,zhang2015magnetic,
luo2018reconfigurable} have been proposed. The skyrmion velocity in these devices is limited by 
its transverse displacement due to Magnus 
force~\cite{fert2013skyrmions}. Prior works have considered 
this effect as a constraint, severely limiting their performance. 
Magnetic bilayer systems, where the skyrmions nucleated in each of these
layers are anti-ferromagnetically coupled to each
other~\cite{zhang2016bilayerskyrmion,woo2018current} such that the 
Magnus force cancels out, have been proposed. However, in such cases, 
careful engineering of the materials is required to obtain perfect 
coupling of the skyrmions. 

In this work, we propose a skyrmion-based logic device, SkyLogic. 
A spin-polarized charge current is injected into a PMA ferromagnet 
(PMA--FM) to nucleate a skyrmion at the 
input end. A charge current sent through a high resistivity material, called 
the spin Hall metal (SHM) placed directly beneath the PMA--FM 
propagates the skyrmion from the input end to the output end of the PMA--FM. 
We counter the transverse displacement of the skyrmion from the Magnus 
effect by sending a charge current through a repeater SHM (R--SHM) placed 
at intervals above the PMA--FM, and perpendicular
to the direction of the SHM. The skyrmion is detected at the output 
end as a resistance change in the magnetic tunnel junction (MTJ) stack. 
A charge current sent through the MTJ stack turns the input transistor 
switch of the next stage SkyLogic on. Depending on the resistance of the 
MTJ stack, the strength of the charge current nucleates a skyrmion in the 
next stage, thereby implementing an inverter. 

We derive an analytical model to analyze the skyrmion motion through
the PMA--FM under the application of the two charge currents. 
We obtain the delay and energy model of SkyLogic as a function of 
the PMA--FM material parameters. Next, we perform a systematic 
design space exploration of the material parameters to optimize the 
device performance. We show that our novel approach to counter 
the Magnus force allows the use of large current densities for 
skyrmion propagation. In addition, our circuit-based
solution can be implemented with the existing materials
that have been used to demonstrate skyrmion propagation.
With the novel design and optimization, we show that it is 
possible to achieve an inverter delay of $434$ps, 
with $7.1$fJ switching energy. 

\section{Overview of the SkyLogic Device}
We first propose the SkyLogic device and explain it using an inverter. 
We then show the design of a two-input NOR (NOR2) gate implemented
using SkyLogic. 

\subsection{Design of the SkyLogic Inverter}
\label{sec:skyinv}
A schematic of our proposed skyrmion--based logic device is shown in
Fig.~\ref{fig:device}~(a), and consists of the following components.  At the input
end, a polarizer ferromagnet (P--FM) is placed on top of ferromagnetic
intra--gate interconnect with perpendicular magnetic anisotropy, PMA--FM,
which connects the input to the output of the device. A layer of SHM is 
placed below the PMA--FM, along its entire length. Another layer of SHM, called the
repeater SHM (R--SHM), is placed on top of the PMA--FM between the input and the
output in a direction transverse to the PMA--FM. At the output end, the
presence of a skyrmion is detected by an MTJ
structure. This consists of an oxide layer sandwiched between a ferromagnet
with pinned magnetization (Pin--FM) and the output end of the PMA--FM channel.
The output end of the PMA--FM acts as the MTJ free layer. The functionality 
of the device can be understood by examining 
its operation in stages:

\noindent
{\bf{Stage 1 -- Skyrmion nucleation:}}
At the input end, a voltage $V_{in}$ turns a transistor $T_{in}$ on and
sends a charge current with current density $J_{nuc}$. This charge
current is spin--polarized by the P--FM~\cite{fert2017review}. If
$J_{nuc}$ is greater than a critical current density
$J_{c,nuc}$, the spin-polarized current nucleates a skyrmion in the 
PMA--FM beneath the P--FM, in time $t_{nuc}$.

\noindent
{\bf{Stage 2 -- Skyrmion propagation:}}
Once the skyrmion is nucleated, a voltage $V_{prop,x}$ turns on transistor
$T_{prop,x}$ and sends a charge current with density $J_x$ through the SHM. As a
result, the skyrmion propagates from the input end to the output end of the
PMA--FM due to the spin Hall effect (SHE) in time $t_{prop}$. As the 
skyrmion is propagated longitudinally along the x--direction, 
it also experiences a transverse motion in the y--direction due to the effect of Magnus
force~\cite{sampaio2013nucleation}. 

\begin{figure}[ht]
    \subfigure[]{
		\includegraphics[width=0.18\textwidth]{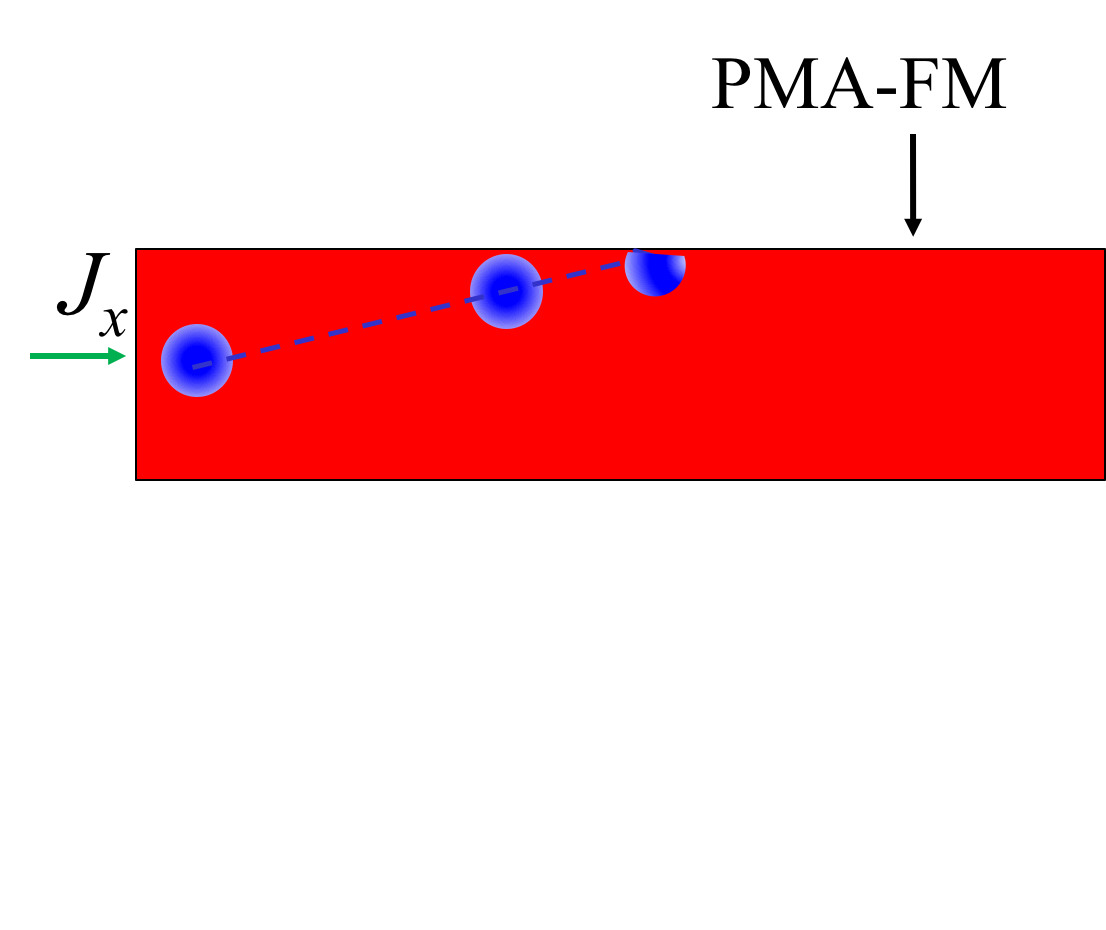}
	}
	 \subfigure[]{
		\includegraphics[width=0.3\textwidth]{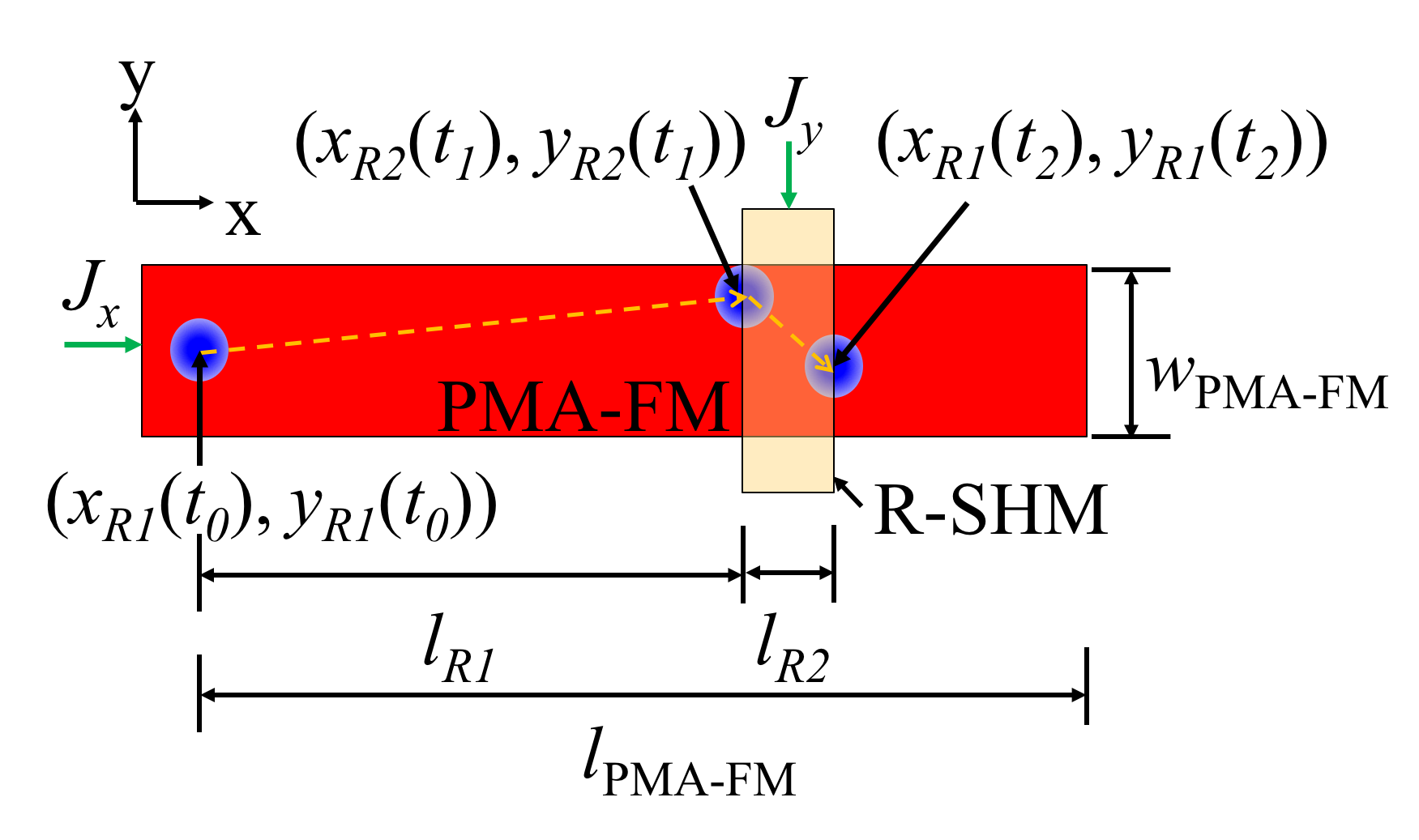}
	}
    \caption{Skyrmion trajectory in the PMA--FM (a) without R-SHM inserted,
		showing annihilation, and (b) with R-SHM inserted, avoiding
		annihilation.}
    \label{fig:repeater}
\end{figure}

When $J_x$ is
applied to the SHM, as shown in Fig.~\ref{fig:repeater}(a), 
the transverse motion of the skyrmion causes it to travel towards the edge of 
the PMA--FM where it is annihilated. 
The placement of R--SHM above the PMA--FM addresses this issue.
As shown in Fig.~\ref{fig:repeater}(b), we assume equally-spaced repeaters
above the PMA--FM, and define segments of the PMA--FM that do not lie under
R--SHM as Region 1 ($R1$), with length $l_{R1}$.  The segments of the PMA--FM
below R--SHM is denoted as Region 2 ($R2$), with length $l_{R2}$.  The
width of the PMA--FM is denoted as $w_{\text{PMA--FM}}$.

The role of the repeater is to deflect the skyrmion back into the body of the
PMA--FM, and $l_{R1}$ is chosen to ensure the deviation due to the Magnus force
does not allow the skyrmion to reach the PMA--FM edge where it would be annihilated.  
The transistor $T_{prop,y}$ is turned on by applying a gate voltage $V_{prop,y}$ 
and a charge current with density $J_y$ is applied through the R--SHM layer 
in a direction transverse 
to that of $J_x$, as shown in Fig.~\ref{fig:repeater}(b). Therefore in $R1$, 
the skyrmion motion is
defined by the forcing function from $J_x$ and in $R2$, the skyrmion motion is
dictated by the forcing function from both $J_x$ and $J_y$.  Due to $J_y$, the
skyrmion experiences a longitudinal motion in the y--direction and a Magnus
force in the x--direction. The magnitudes of $J_x$ and $J_y$ can be optimized such that the
skyrmion moves back towards the PMA--FM interior from its edge to negate
the y--direction shift due to the Magnus force from $J_x$.  At the same time,
the Magnus force due to $J_y$ moves the skyrmion forward towards the output.

\begin{figure}[ht]
	\centering
	\includegraphics[width=7cm]{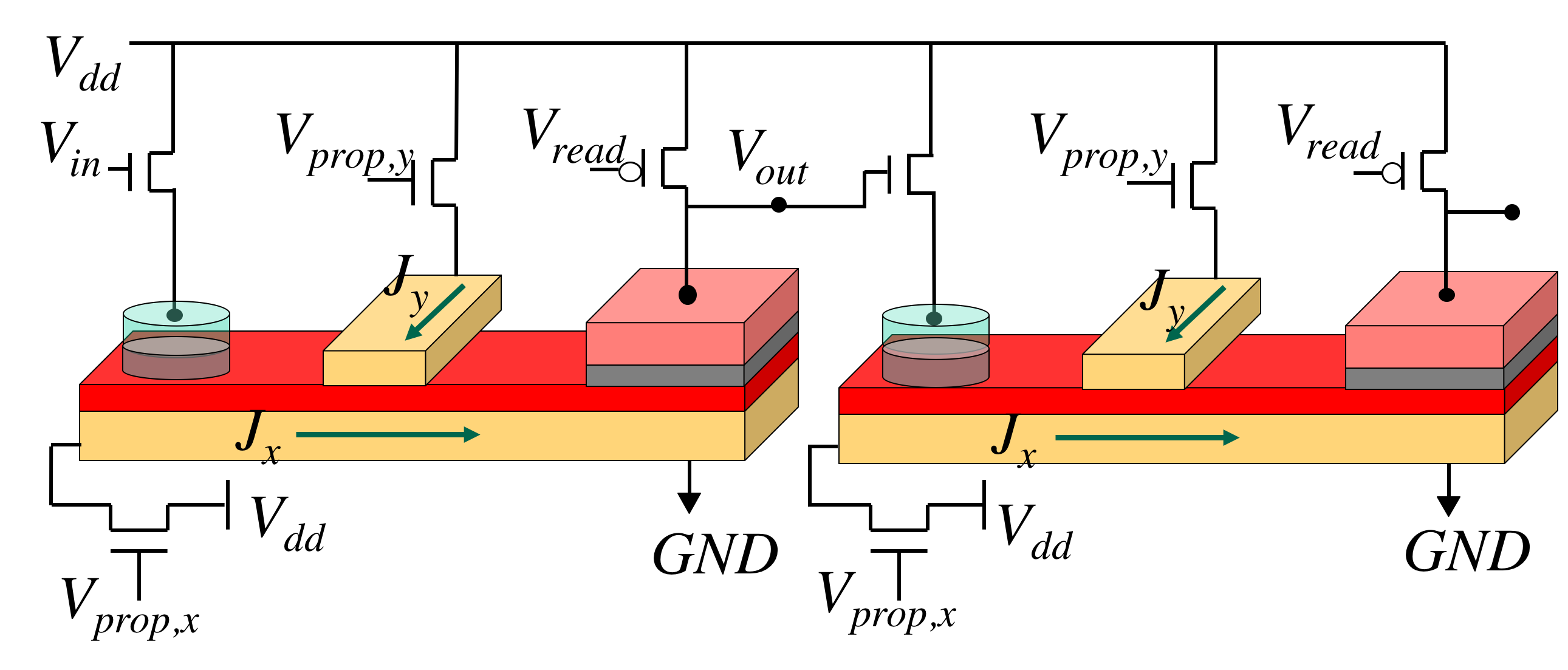}
	\caption{A pair of cascaded SkyLogic devices, indicating how the
		output of one device is transferred to the input of the next stage.}
	\label{fig:cascade}
\end{figure}

Additional repeaters can be inserted along the PMA--FM length to allow
the skyrmion to traverse long interconnects. Our approach with
repeaters differs from conventional
approaches~\cite{sampaio2013nucleation,jiang2017direct,RaposoSkyrmion,
tomasello2014strategy,woo2017spin} that only use $J_x$ applied at one end of
the SHM to propagate the skyrmion in the PMA--FM. In these approaches, a low
current density is essential to contain the skyrmion within the
PMA--FM, directly translating to high skyrmion propagation delays. In contrast, 
our novel approach of inserting R--SHM and using both $J_x$ and $J_y$ allows the use of
high current densities for fast propagation of skyrmions through the PMA--FM,
while ensuring that the skyrmion is not annihilated due to its transverse
motion. 

\noindent
{\bf{Stage 3 -- Skyrmion detection:}}
Once the skyrmion reaches the output end, it creates a polarization in the free
layer of the MTJ stack at the output. The Pin--FM magnetization is
anti-parallel to that of the steady-state PMA--FM. 
The presence or absence of a skyrmion is differentiated by different 
resistances for the MTJ stack for these two cases~\cite{hanneken2015electrical,
tomasello2017electrical}: the
resistance is high if no skyrmion is present, and low otherwise.
The time required for the skyrmion detection is denoted by $t_{det}$.

\noindent
{\bf {Stage 4 -- Cascading logic stages:}}
The cascading of successive SkyLogic stages can be achieved as shown in
Fig.~\ref{fig:cascade}. The voltage $V_{read}$ is set to low to turn on the
transistor $T_{read}$ and a voltage $V_{out}$ is induced at the output node. 
This switches the transistor $T_{in}$ in the next logic stage on. 
The current density through $T_{in}$ is greater than the critical nucleation 
current density, $J_{c,nuc}$, when the skyrmion is
absent at the output end in the PMA--FM layer of the MTJ. 
Therefore, a skyrmion is nucleated; when the current density is lower than
the critical value, the skyrmion is present at the output, and no nucleation
occurs at the next stage. Thus, we realize an inverter.

\begin{figure}[ht]
	\centering
\subfigure[] {
	\includegraphics[width=7cm]{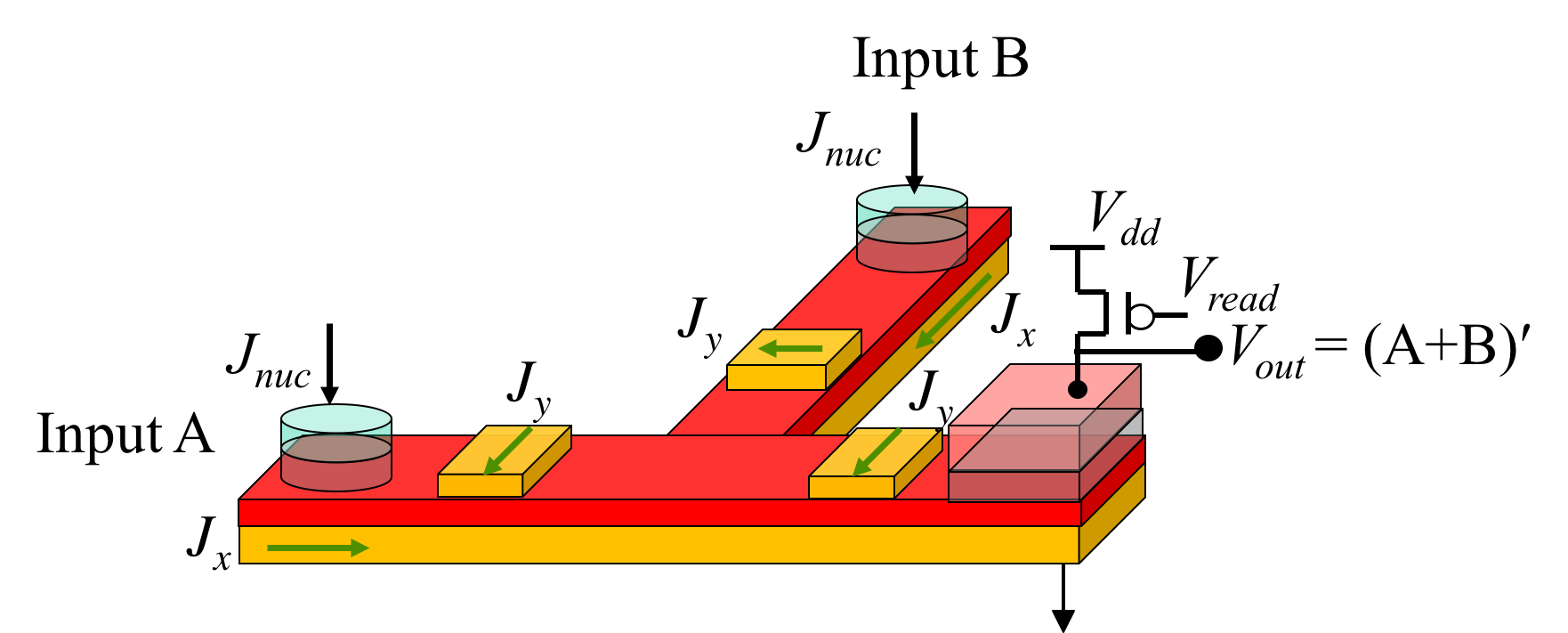}
}
\caption{ (a) The design of a SkyLogic NOR2 gate.}
\label{fig:skynor}
\end{figure}

\subsection{Design of the SkyLogic NOR gate}
\label{sec:norgate}
The presence of a skyrmion in at least one of the input
results in a skyrmion being absent (present) at the output to realize a 
(N)OR gate. Skyrmion-based 
(N)OR gates has been shown in~\cite{zhang2015magnetic}. 
We use this design and combine it with that of the 
SkyLogic inverter presented in Section~\ref{sec:skyinv} to build a NOR
gate, as shown in Fig.~\ref{fig:skynor}. For clarity, we do not
show the CMOS transistors. 

The SkyLogic NOR2 gate consists of two input branches and an 
output branch. For each of the two inputs, when a current with density 
$J_{nuc}$ is applied, a skyrmion is nucleated if $J_{nuc} > J_{c,nuc}$, 
thereby representing logic 1. The absence 
of a skyrmion, when $J_{nuc} < J_{c,nuc}$, represents logic 0. 
A charge current with density $J_x$ sent through the SHM through 
each of the input branches propagates the skyrmion from the 
input branch towards the output branch of the PMA--FM. As in the case 
of the SkyLogic inverter, R--SHMs are inserted in the input branches 
atop the PMA--FM to correct the course of the skyrmion. 
At the output end, if any one of the inputs is logic 1, i.e., 
if at least one skyrmion is nucleated at the input end, then a skyrmion 
is present at the output end. The skyrmion is then detected using 
the MTJ stack, as explained in Stage 3 of Section~\ref{sec:skyinv}. 
The voltage at the output node, $V_{out}$, therefore represents 
a logical NOR of the two inputs. 

\section{Performance modeling of SkyLogic inverter}
\label{sec:modeling} 

In this section, we analyze the performance of SkyLogic 
inverter shown in Fig.~\ref{fig:device} in the four stages of 
its operation.

\subsection{Skyrmion nucleation}
We model the skyrmion nucleation using the micromagnetic simulator 
OOMMF~\cite{oommf} with the DMI extension~\cite{rohart2013skyrmion}. 
The PMA--FM is initially uniformly magnetized in the $+$z direction. 
A spin-polarized current is sent through the P--FM in the $-$z direction. 
If the current density 
through the P--FM, $J_{nuc}$, is greater than $J_{c,nuc}$, then a
skyrmion is nucleated in the PMA--FM layer beneath the P--FM.

\subsection{Skyrmion propagation}
\label{sec:skypropmodel}

Next, we separately analyze the propagation of the skyrmion in regions
$R1$ and $R2$ using Fig.~\ref{fig:repeater}(b). 

\noindent
{\bf Case 1: Skyrmion propagation in $R1$:}
The motion of the skyrmion in $R1$ can be explained using 
the two-dimensional Thiele equation~\cite{ThieleOriginal,
tomasello2014strategy,RaposoSkyrmion,jiang2017direct,sampaio2013nucleation} 
as follows:
\begin{equation}
	\vec{G} \times \vec{v}_{R1} + \alpha \vec{D} \cdot \vec{v}_{R1} =
\vec{F}_{SHE,x} +
	\vec{F}_{c,y} \label{eq:ThieleRegion1} \\ 
\end{equation}

\noindent
where $\vec{G} = \{0, 0, G\} = \Big\{0,0,\frac{-4{\pi}Q{M_s}
t_{\text{PMA--FM}}}{\gamma_0}\Big\}$ is the gyrovector, $M_s$ is the PMA--FM saturation
magnetization, $t_{\text{PMA--FM}}$ is the PMA--FM thickness, $\gamma_0$ is the 
gyromagnetic ratio, $Q = +1$/$-1$ is the skyrmion chirality, and
$k$ is the confinement constant. The skyrmion velocity in $R1$ is denoted
by its x-- and y-- components as $\vec{v}_{R1} = \{v_{x,R1}, v_{y,R1}\} =
\Big\{\frac{d}{dt}(x_{R1}(t)),\frac{d}{dt}(y_{R1}(t))\Big\} $. 
The time-dependent x-- (y--) position of the skyrmion center
in $R1$ of the PMA--FM is given by
 $x_{R1}(t)$ ($y_{R1}(t)$). The skyrmion driving force due to the SHE 
as a result of $J_x$, is given by $\vec{F}_{SHE,x} = \{F_{SHE,x},0\} = 
\Big\{\frac{\hbar \theta_{SHE} J_x Q \pi^2 R_{sk}
}{2e},0\Big\}$. The damping constant 
is denoted by $\alpha$ while the dissipative force tensor is given by $\vec{D}=
\begin{bmatrix} 
	D & 0 \\
 	0 & D 
\end{bmatrix}
; D = \Big\{\frac{-M_st_{\text{PMA--FM}}\pi^3R_{sk}}{\Delta
\gamma_0}\Big\}$. 
The skyrmion radius is given by $R_{sk}$, $\Delta$ denotes its domain-wall
width and $\theta_{SHE}$ represents the spin Hall angle. The
repelling force experienced by the skyrmion from the PMA--FM edges along
its width is given by $\vec{F}_{c,y} = \{0,F_{c,y}\} =  \{0,
-ky_{R1}(t)\}$.
 
The first term in Equation~\eqref{eq:ThieleRegion1},
$\vec{G}\times\vec{v}_{R1}$, represents the impact of the Magnus force on the
skyrmion, and defines its transverse motion. The second term in
Equation~\eqref{eq:ThieleRegion1}, $\alpha\vec{D}\cdot\vec{v}_{R1}$, is the
opposing force experienced by the skyrmion due to the intrinsic damping of the
PMA--FM. At steady state, these forces are countered by the driving force due
to SHE, $\vec{F}_{SHE,x}$, and the repelling force, $\vec{F}_{c,y}$. We
solve the 2D first-order differential equation~\eqref{eq:ThieleRegion1} to obtain
$x_{R1}(t)$ and $y_{R1}(t)$:
\begin{align*}
	x_{R1}(t) &= x_{R1}(t_0) + \frac{t}{\tau}\Bigg[\frac{F_{SHE,x}}{k}
- \frac{G}{\alpha D} y_{R1}(t) \Bigg] \numberthis \label{eq:R1XDisp} \\
	y_{R1}(t) &= \frac{G F_{SHE,x}}{\alpha D k} \Big(e^{-t/\tau}  
				- 1 \Big) + y_{R1}(t_0)e^{-t/\tau}  \numberthis \label{eq:R1YDisp} 
\end{align*}
\noindent 
where $\tau = \left|{\frac{G^2 + (\alpha D)^2}{\alpha D k}}\right|$
is the characteristic relaxation time of the skyrmion. At time $t_0$, (x,y) co-ordinates
of the center of the nucleated skyrmion is given by ($x_{R1}(t_0)$,$y_{R1}(t_0)$). 
The term $\frac{G}{\alpha D}$ represents the ratio of the Magnus force and 
dissipative force. The relative magnitude of each of these forces determines
the net strength of the opposing force to the skyrmion propagation.  
The $\vec{F}_{c,y}$ term
does not have an $\vec{F}_{c,x}$ counterpart because
there is no repelling force on the skyrmion from the PMA--FM edges along its
length. 

\noindent {\bf Case 2: Skyrmion propagation in $R2$:} We modify
Equation~\eqref{eq:ThieleRegion1} to model the skyrmion motion in $R2$, in
which both $J_x$ and $J_y$ are active, by adding $\vec{F}_{SHE,y}$ to
the right hand side. Here, $\vec{F}_{SHE,y} = \{0, F_{SHE,y}\} = \Bigg\{0,\frac{ \hbar\theta_{SHE}
J_y Q \pi^2 R_{sk}}{2e}\Bigg\}$ is the force experienced by the skyrmion as a
result of the SHE arising from $J_y$ in Region 2.  
The instantaneous skyrmion velocity in $R2$ is given by $\vec{v}_{R2} =
\{v_{x,R2},v_{y,R2}\}$ with $x_{R2}(t)$
($y_{R2}(t)$) denoting the time-dependent x--position (y--position) of the
center of the skyrmion in $R2$. We solve the 2D first-order differential equation
for $R2$ to obtain $x_{R2}(t)$ and $y_{R2}(t)$:
\begin{align*}
	x_{R2}(t) =&  x_{R2}(t_1) + \frac{t}{\tau}\Bigg[\frac{F_{SHE,x}}{k} +
\frac{GF_{SHE,y}}{\alpha D k} - 
	\frac{G}{\alpha D}y_{R2}(t)\Bigg]
\numberthis \label{eq:R2XDisp} \\
	y_{R2}(t) =&  y_{R2}(t_1) e^{-t/\tau} + \Bigg[\frac{GF_{SHE,x}}{\alpha D k} -
\frac{F_{SHE,y}}{k}\Bigg]\Big( e^{-t/\tau} - 1
\Big) \numberthis 
\label{eq:R2YDisp}
\end{align*}
At time $t_1$, the skyrmion reaches the PMA--FM edge in $R1$ and a current 
$J_y$ is applied along R-SHM length as shown in Fig.~\ref{fig:repeater}(b). 
The (x,y) coordinates of the skyrmion center at time $t_1$ is given by 
($x_{R2}(t_1)$, $y_{R2}(t_1)$). Equation~\eqref{eq:R2XDisp} shows that the 
skyrmion is propelled forward towards the PMA--FM output in $R2$ by 
the combined force of $F_{SHE,x}$ and the Magnus force resulting
from applying $J_y$ in R-SHM. The value of $J_y$ in $R2$ should be 
chosen such that the skyrmion continues to move back towards the PMA--FM
interior upon its application. This can be achieved by enforcing
the constraint $x_{R2}(t) > x_{R2}(t_1)$ and 
$y_{R2}(t) < y_{R2}(t_1)$ in Equations~\eqref{eq:R2XDisp}
and~\eqref{eq:R2YDisp}. 

For a fixed PMA--FM length, the number of required R-SHMs, $p$, depend on the choice 
of $J_x$ and $J_y$ and can be determined by solving the skyrmion displacement 
equations for each of $R1$ and $R2$ regions independently. The average
net velocity (propagation time) of the skyrmion, $v_x$ ($t_{prop}$), 
as it propagates from the PMA--FM input to its output with $p$ repeaters,  
is therefore given by the sum of their velocities (propagation times) in $R1$ and $R2$. 

\begin{figure}[ht]
	\subfigure[]{
		\centering
		\includegraphics[width=3.5cm]{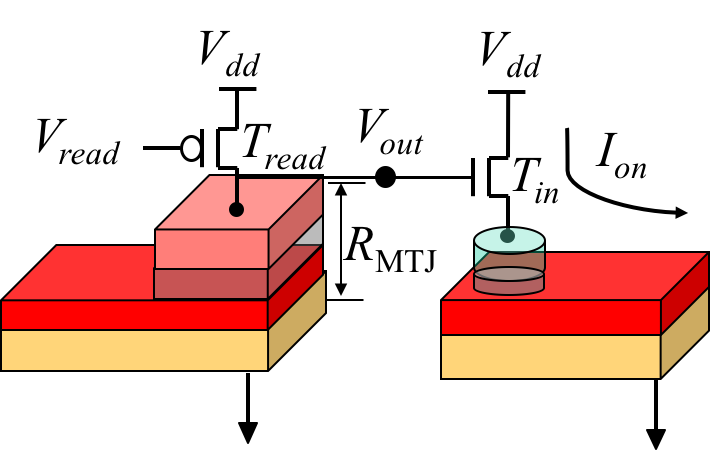}
	}
	\subfigure[]{
		\centering
		\includegraphics[width=3.5cm]{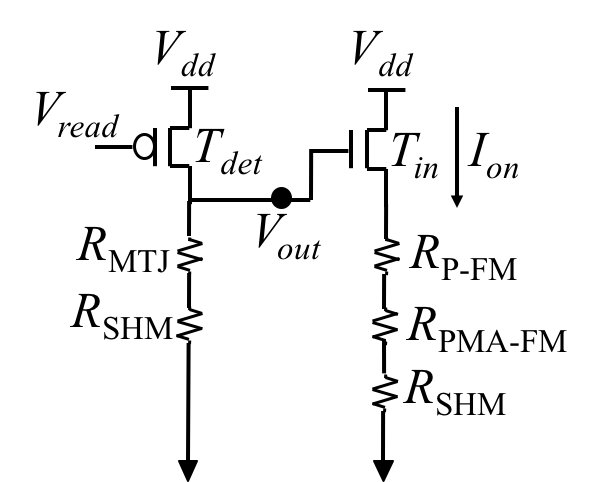}
	}
\caption{(a) Two cascading SkyLogic devices (b) The equivalent circuit.}
\label{fig:eqDetectionCkt}
\end{figure}

\subsection{Skyrmion detection and cascading SkyLogic devices } 
We explain the detection of the skyrmion using the MTJ stack. 
The initial magnetization in the Pin--FM layer is uniform in $-z$ direction, 
whereas that of the PMA--FM layer is in the $+$z direction.    
Let $R_{\text{MTJ,0}}$ ($R_{\text{MTJ,1}}$) denote the 
resistance of the MTJ stack when the skyrmion is absent (present) 
in the PMA--FM layer, which forms the free layer of the MTJ stack. 
We can write $R_{\text{MTJ,0}}$ and $R_{\text{MTJ,1}}$ as
\begin{align}
	R_{\text{MTJ,0}} & \ = \ R_{AP}; R_{\text{MTJ,1}} = \eta R_{AP} + (1-\eta) R_P.\label{eq:rmtj1}
\end{align}
\noindent
Here, $R_{P}$ ($R_{AP}$) corresponds to the resistance of the MTJ stack
when its pinned layer and the free layer are parallel (anti-parallel) to
each other. In the absence of the skyrmion (logic 0) in the PMA--FM layer, 
 $R_{\text{MTJ,0}}$ corresponds to $R_{AP}$. However, when
the skyrmion is present in the PMA--FM layer (logic 1), the
PMA--FM layer magnetization is not completely parallel to
the pinned layer due to the averaging nature of the magnetization profile 
of the skyrmion~\cite{kang2016voltage,zhang2015SkyrmionTx,
weishangSkyrmionSpectrum}. 
We therefore model $R_{\text{MTJ,1}}$ as a linear combination of $R_{P}$ and 
$R_{AP}$ with $\eta$ being the scalar coefficient defined as 
$\eta = \frac{A_{skyrmion}}{A_{det}}$. Here $A_{skyrmion}$ refers 
to the area of the skyrmion while $A_{det}$ 
refers to the area of the detector enclosed by the length of the fixed
layer of the MTJ, $l_{det}$, and its width, $w_{det}$. 
Since $A_{skyrmion} < A_{det}$, it follows
that $\eta < 1$. 

We analyze the cascading of two SkyLogic devices, shown in
Fig.~\ref{fig:eqDetectionCkt}(a), with the equivalent circuit shown in
Fig.~\ref{fig:eqDetectionCkt}(b). The voltage at the output node,
$V_{out}$ is determined by the voltage divider circuit formed by the
resistances of transistor $T_{read}$ ($R_{TX}$), MTJ
($R_{\text{MTJ}}$), and SHM ($R_{\text{SHM}}$).  
Once the transistor $T_{read}$ is turned on, the current in the
transistor $T_{in}$ in the next SkyLogic stage, $I_{on}$, is
proportional to $V_{out}$ and its strength determines whether a 
skyrmion is nucleated at the next stage. 

\subsection{Modeling performance}
\label{sec:perfmodel}

Here, we outline the model used to measure the performance of
SkyLogic inverter. We model the delay, $T_{\text{SkyLogic}}$, and energy,
$E_{\text{SkyLogic}}$, of the SkyLogic inverter in
Fig.~\ref{fig:device} as follows.
\begin{equation}
	\begin{aligned}
		T_{\text{SkyLogic}} &= t_{nuc} + t_{prop} +
t_{det} \\
		E_{\text{SkyLogic}} &= E_{nuc} + E_{prop} +
E_{det} + E_{TX}.
	\end{aligned}
\label{eq:performance}
\end{equation}
\noindent
The energy dissipated during the nucleation, propagation, detection, and
peripheral CMOS transistor switching is given respectively by $E_{nuc}$,
$E_{prop}$, $E_{det}$, and $E_{TX}$. The energy terms $E_{nuc}$ and
$E_{det}$ represent the Joule heating during the nucleation and
detection process, while $E_{prop}$ represents both Joule heating due to
$J_x$ and $J_y$, and the energy required to turn on the CMOS transistors
that supply both $J_x$ and $J_y$. The energy required to turn on the
rest of the CMOS transistors is grouped in the term $E_{TX}$.

\begin{table}[hb]
     \centering
     \caption{Simulation parameters used in this work.}
	\label{tbl:parameters}
	\footnotesize
     \begin{tabular}{|l|r|}
         \hline
         {\bf Parameter}                                                                & {\bf Value}                \\ \hline
		 $l_{\text{PMA--FM}}{\times}w_{\text{PMA--FM}}{\times}h_{\text{PMA--FM}}$ [nm$^3$] & 200$\times$50$\times$0.4 \\ \hline
		  $l_{\text{SHM}}{\times}w_{\text{SHM}}{\times}h_{\text{SHM}}$ [nm$^3$]& 200$\times$50$\times$1  	 \\ \hline
		 $l_{\text{R--SHM}}{\times}w_{\text{R--SHM}}{\times}h_{\text{R--SHM}}$
[nm$^3$]& 25$\times$50$\times$1  	 \\ \hline
         Exchange constant, $A$ [pJ/m]                                                  & 15  					     \\ \hline
		 Radius of the skyrmion, $R_{sk}$ [m]									        & $8 \times 10^{-9}$         \\ \hline
		 Confinement constant, $k$ [N/m]											    & $-3.6 \times 10^{-5}$	 \\ \hline 
          $\rho_{\text{SHM}}$, $\rho_{\text{R--SHM}}$ {[}$\Omega$-m{]}				        & $1.06 \times 10^{-7}$      \\ \hline
		 $\rho_{\text{PMA--FM}}$ [$\Omega$-m]
& $1.7 \times 10^{-7}$ 							 \\ \hline
         Spin Hall angle, $\theta_{\text{SHE}}$											& 0.33                       \\ \hline
         Spin polarization, $P_{\text{P--FM}}$										& 1                 	     \\ \hline
		TMR of the output MTJ stack [$\%$] & 300 \\ \hline
		$R_{AP} = R_{\text{MTJ,0}}$ [$\Omega$] & 4000 \\ \hline 
		$\eta$  & 0.5 \\ \hline
		CMOS transistor gate capacitance, $C_g$ [F] & $0.1\times10^{-15}$ \\ \hline
		$V_{dd}$, $V_{read}$ [V] & 1 \\ \hline
		$V_{prop,x}$, $V_{prop,y}$ [V] & 0.25 \\ \hline
    \end{tabular}
\end{table}

\section{Results and Discussion}

In this section, we demonstrate the SkyLogic device design and obtain 
its delay and energy with the help of the models explained in 
Section~\ref{sec:modeling} and the
peripheral CMOS circuity implemented in the 10nm Predictive Technology
Model (PTM)~\cite{PTM} with the parameters shown in Table~\ref{tbl:parameters}. 

\subsection{Insertion of R--SHM}
\label{sec:rshm}
\begin{figure}[hb]
\centering
\subfigure[]{
	\includegraphics[width=3.8cm]{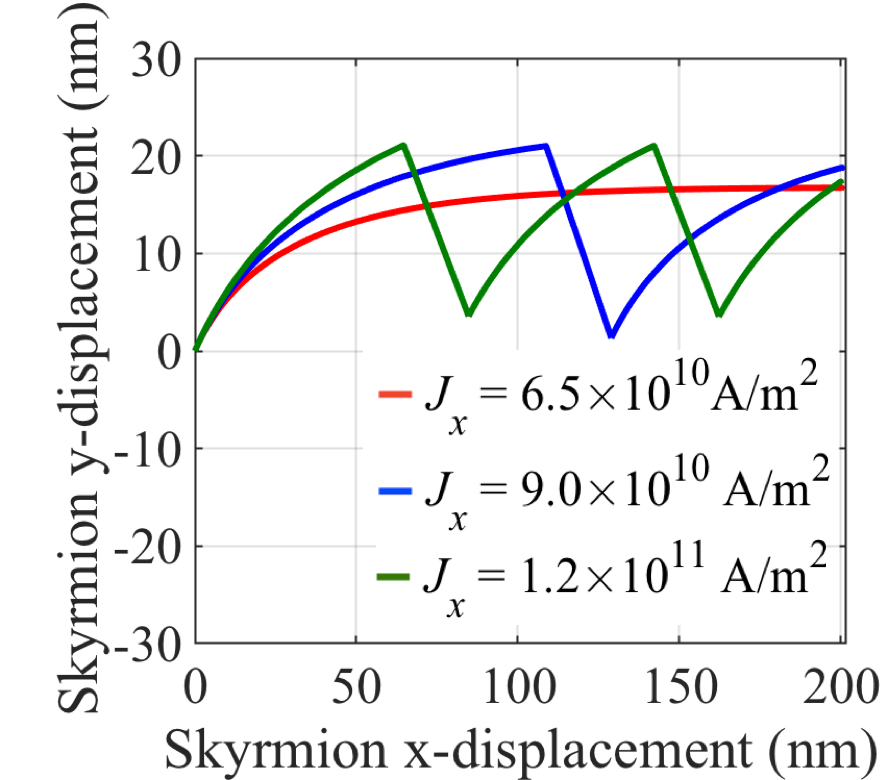}
}
\subfigure[]{
	\centering
	\includegraphics[width=4.4cm,height=3.1cm]{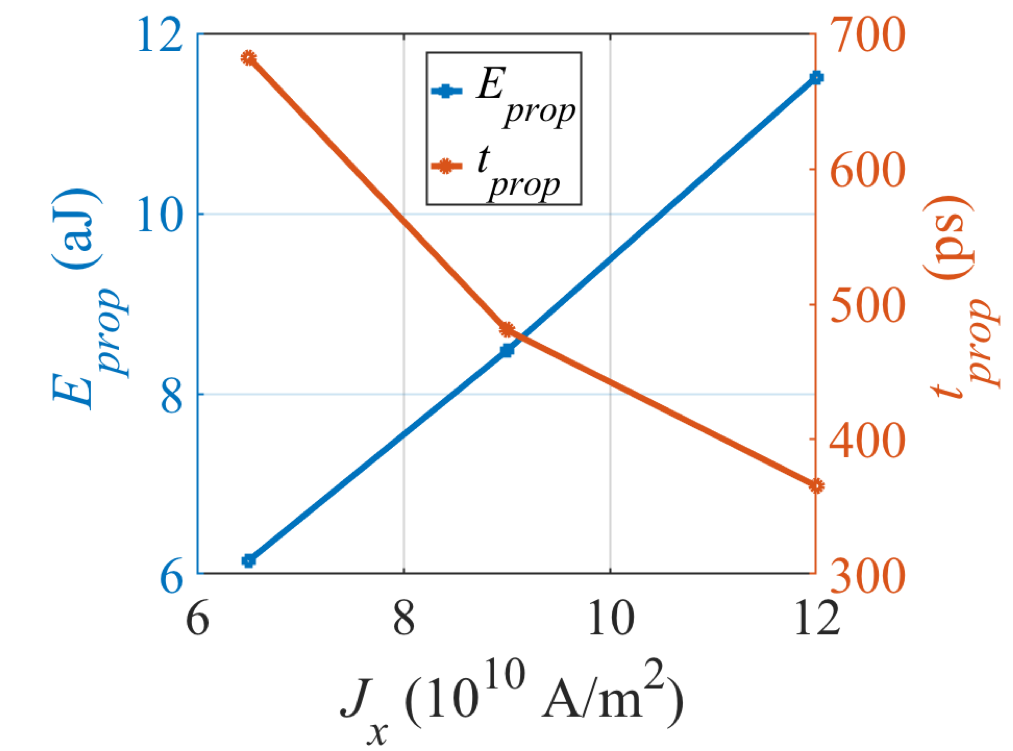}
}
\caption{(a) Skyrmion trajectory in the PMA--FM for various values of
$J_x$. (b) Skyrmion propagation energy, $E_{prop}$, and propagation
delay, $t_{prop}$, as a function of $J_x$. The material
parameters used in this simulation are: $M_{S,\text{PMA--FM}} =
1\times10^5$ A/m, $K_{U,\text{PMA--FM}} = 8\times10^5$ J/m$^3$, and
$\alpha = 0.25$.}
\label{fig:rshmInsertion}
\end{figure}

Here, we examine the process of R--SHM insertion in a SkyLogic device. 
We use the following PMA--FM material parameters in our simulations:
$M_{S,\text{PMA--FM}} = 1\times 10^5$ A/m, $K_{U,\text{PMA--FM}} =
8\times 10^5$ J/m$^3$, $\alpha = 0.25$. We show, at the end of
Section~\ref{sec:skyprop}, that this design point gives the best
energy-delay product for skyrmion propagation phase. 
We assume that the center of the nucleated skyrmion is the origin of the 
coordinate system. For various values of $J_x$, we simulate the skyrmion trajectory by 
solving the displacement equations shown in Section~\ref{sec:skypropmodel} 
and plot the results in Fig.~\ref{fig:rshmInsertion}(a). 

For $J_x = 6.5\times 10^{10}$ A/m$^2$, the skyrmion reaches the PMA--FM 
output without reaching the edge due to low transverse and longitudinal 
velocity. Next, when $J_x$ is increased to $9\times10^{10}$ A/m$^2$, 
$v_{y,R1}$ is greater than $v_{x,R1}$ and 
therefore reaches the PMA--FM edge faster than the
case where $J_x = 6.5\times10^{10}$ A/m$^2$. In this case, an R--SHM needs
to be inserted to avoid the skyrmion being annihilated at the edge. When
an R--SHM is inserted and $J_y = 5\times10^{11}$ A/m$^2$ is applied 
through it, the skyrmion is pushed back into the interior 
of the PMA--FM in $R2$ underneath the R--SHM. When $J_x$ is further 
increased to $1.2\times10^{11}$ A/m$^2$, $v_{y,R1}$ further increases 
relative to increase in $v_{x,R1}$ and causes the skyrmion to reach the 
PMA--FM edge faster than the earlier two cases. Two R--SHMs are required 
in this case between the PMA--FM input and output. 

Next, we calculate the cost of inserting an R--SHM on $t_{prop}$ and 
$E_{prop}$ and plot the results in Fig.~\ref{fig:rshmInsertion}(b). 
As $J_x$ increases, $t_{prop}$ decreases because of an increase in
$v_{x,R1}$ and also due to the effect of the Magnus force in $R2$ along the
x--direction due to $J_y$ which propels the skyrmion faster towards the output. 
The propagation energy, $E_{prop}$, however increases.
Though $E_{prop}$ is directly proportional to $t_{prop}$, the R--SHM 
insertion requires an additional transistor to drive $J_y$ through 
it, thereby increasing $E_{prop}$. The net energy-delay product, $EDP_{prop}$, as 
$J_x$ is increased from $6.5 \times 10^{10}$ A/m$^2$ to $9\times10^{10}$ A/m$^2$, 
decreases because of the decrease
in $t_{prop}$. With a further increase in $J_x$ to $1.2\times10^{11}$ A/m$^2$, 
$t_{prop}$ decreases, but $E_{prop}$ increases due to
Joule heating from the currents $J_x$ and $J_y$ in SHM and R--SHM,
respectively, and the energy required to turn two 
transistors on. The decrease in $t_{prop}$ is outweighed by the increase
in $E_{prop}$ and therefore $EDP_{prop}$ increases. Therefore,
inserting one R--SHM provides an optimal $EDP_{prop}$ for this design
point for the chosen values of $J_x$ and $J_y$.

\subsection{Skyrmion propagation}
\label{sec:skyprop}
The initial steady-state magnetization of the PMA--FM is set to
$+$z direction. We evaluate $t_{prop}$ and $E_{prop}$ at the design points 
formed by the combination of the following material parameters and their values:
$M_{S,\text{PMA--FM}}$ $\in$ \{$1 \times 10^5$ A/m, $3 \times 10^5$
A/m, $5 \times 10^5$ A/m, $8 \times 10^5$ A/m , $10 \times 10^5$
A/m\}, $K_{u,\text{PMA--FM}}$ $\in$ \{$5 \times 10^5$ J/m$^3$, 
$8 \times 10^5$ J/m$^3$, $10 \times 10^5$ J/m$^3$\}, $\alpha$ $\in$ 
\{0.05, 0.1, 0.15, 0.2, 0.25\}. 
At these design points, we assume that a skyrmion can be
nucleated at the input end of the PMA--FM by injecting a spin polarized
current through P--FM. We study the impact of these parameters on
the skyrmion velocity, device energy, and energy-delay product (EDP).

\begin{figure}[ht]
\centering
	 \subfigure[] {
     	\includegraphics[width=7.5cm]{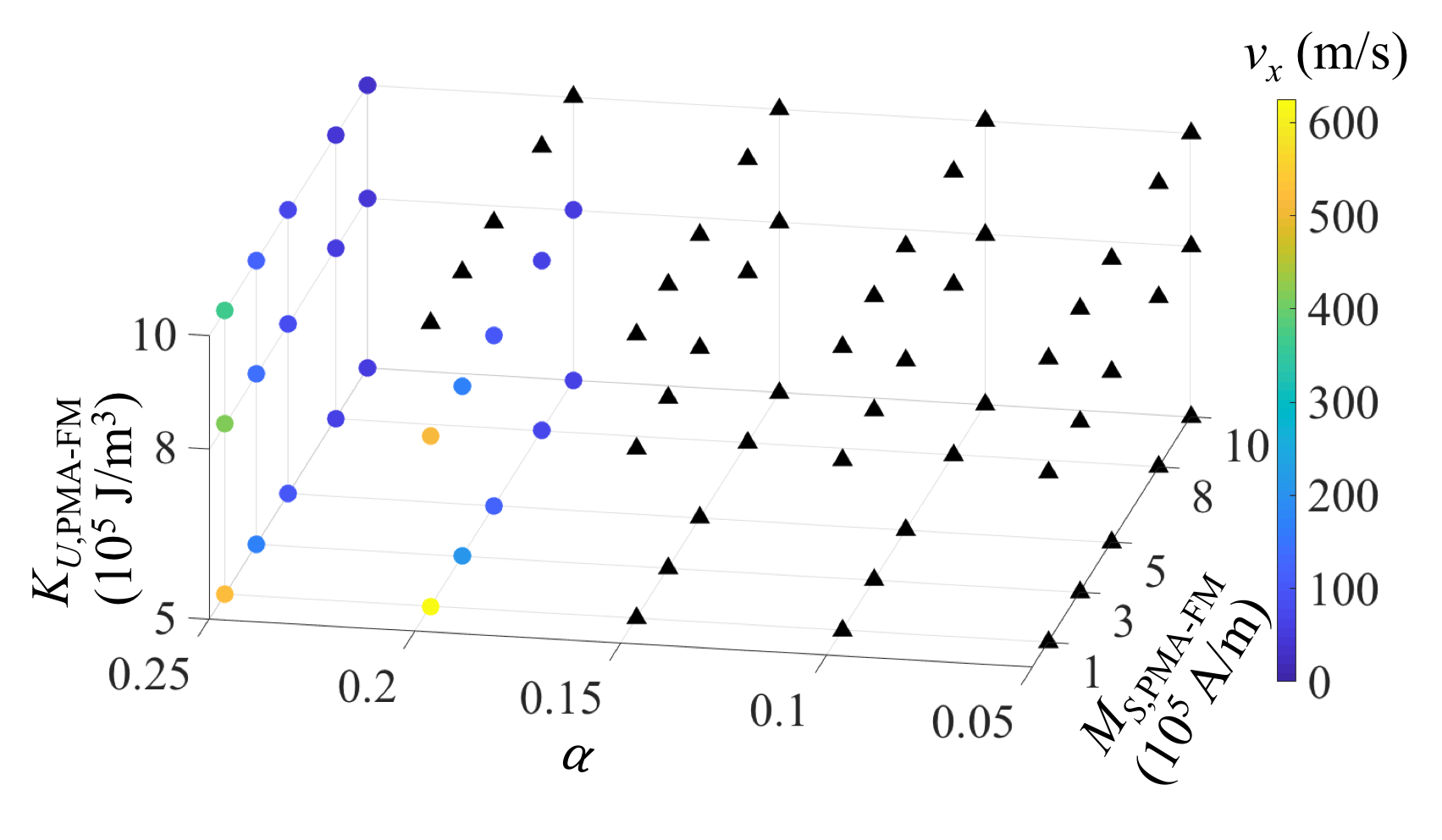}
	}
	\subfigure[]{
		\includegraphics[width=7.5cm]{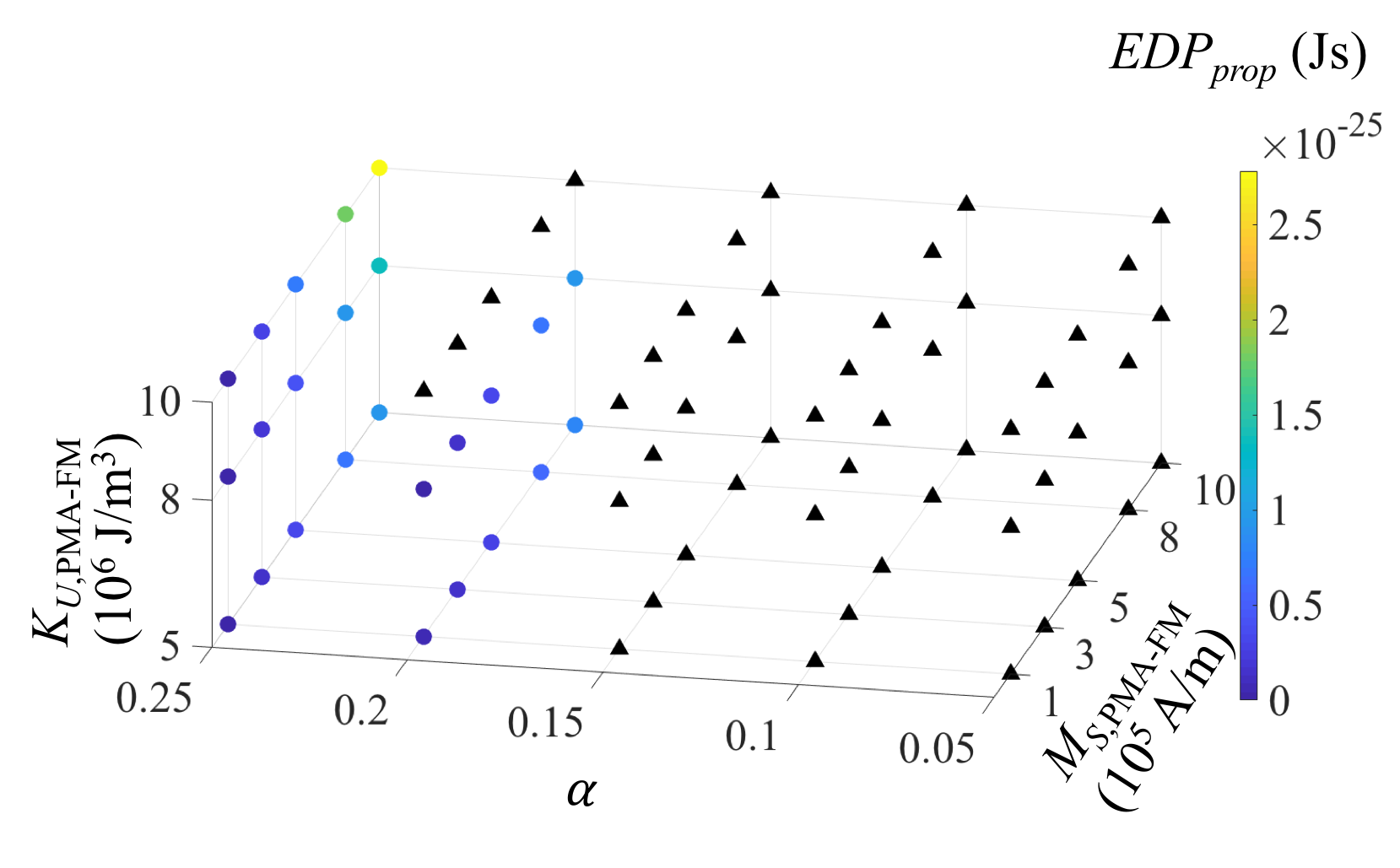}
	}
    \caption{For $J_x = 9 \times 10^{10}$A/m$^2$ and $J_y = 5 \times
10^{11}$A/m$^2$
 (a) net velocity of the skyrmion, $v_{x}$, (b) energy-delay product 
during the propagation of the skyrmion, $EDP_{prop}$.} 
\label{fig:DSEprop}
\end{figure}

\noindent
{\bf Skyrmion velocity, device energy, and EDP:}
We choose $J_x = 9 \times 10^{10}$A/m$^2$ such that 
we obtain $t_{prop} < 500$ ps. We fix the value of $J_y = 5 \times 10^{11}$A/m$^2$.
This choice of $J_y$ ensures that we obtain a valid skyrmion trajectory 
in $R1$ and $R2$ for each design point. We set $p \le 2$, i.e., we
restrict the number of R--SHMs that can be inserted to two 
and disregard the material parameters that violate this criteria. A large 
number of R--SHMs is impractical because additional R--SHMs would incur
extra energy to route $J_y$ through the addition of access transistors leading to higher
$EDP_{prop}$ as we observed in Section~\ref{sec:rshm}. We also disregard 
material parameters where it is impractical to insert a R--SHM in cases
where it would physically overlap with the MTJ structure.
We show $v_x$ and $EDP_{prop}$ as a function of the three material
parameters in Figs.~\ref{fig:DSEprop}(a) and~\ref{fig:DSEprop}(b),
respectively. In these plots, we denote the infeasible design points 
by black triangles.

\noindent
\underline{\em Impact of $\alpha$ and $K_{U,\text{PMA--FM}}$}: 
For $\alpha \le 0.15$, the skyrmion reaches the PMA--FM edge 
along its width faster. This is because for low values of $\alpha$, we
have $v_{y,R1} > v_{x,R1}$ and $v_{y,R2} > v_{x,R2}$, the
skyrmion experiences lower opposition to motion and requires 
more than two R--SHMs, as indicated by the black triangles in 
Fig.~\ref{fig:DSEprop}(a). For $\alpha =0.2$ and 
$K_{U,\text{PMA--FM}} = 10\times10^5$ J/m$^3$, the
skyrmion trajectory is such that it is infeasible to insert an R--SHM.
Therefore, we discard these design points. In general, for a chosen value of
$M_{S,\text{PMA--FM}}$ and $K_{U,\text{PMA--FM}}$, increasing $\alpha$ 
decreases $v_x$, and therefore increases $t_{prop}$ as shown in
Fig.~\ref{fig:DSEprop}(a). The energy-delay product, $EDP_{prop}$, 
shows a corresponding increase, as seen from Fig.~\ref{fig:DSEprop}(b), 
because $E_{prop}$ is directly proportional to $t_{prop}$. 
Hereafter, we restrict our analysis to feasible design points 
that are represented by circles in Figs.~\ref{fig:DSEprop}(a) and~\ref{fig:DSEprop}(b).

As $K_{U,\text{PMA--FM}}$ increases,
the domain-wall width, $\Delta$, decreases by $\sqrt{K_{U,\text{PMA--FM}}}$.
Since the dissipative tensor, $D$, is inversely proportional to $\Delta$, it
follows that increasing $K_{U,\text{PMA--FM}}$ will increase $D$. As the term
${\alpha}D$ determines the strength of the opposition to the skyrmion
propagation, increasing $K_{U,\text{PMA--FM}}$ has the same effect as
increasing $\alpha$. 

\noindent
\underline{\em Impact of $M_{S,\text{PMA--FM}}$}:  
As $M_{S,\text{PMA--FM}}$ increases,
for a given $\alpha$ and $K_{U,\text{PMA--FM}}$, $v_{x}$ decreases as seen in
Fig.~\ref{fig:DSEprop}(a). The dependence of the $R1$ and $R2$ skyrmion
displacement equations~\eqref{eq:R1XDisp} and~\eqref{eq:R2XDisp} on
$M_{S,\text{PMA--FM}}$ can be seen in the term $\tau$. 
Simplifying $\tau$, we determine that $\tau$ is
directly proportional to $M_{S,\text{PMA--FM}}$.  Therefore, for larger
$M_{S,\text{PMA--FM}}$, the skyrmion takes longer to reach the PMA--FM output
because of the larger relaxation time, thereby reducing $v_x$. 
The propagation energy, $E_{prop}$, which is directly proportional to
$t_{prop}$ also increases with increase in $M_{S,\text{PMA--FM}}$, 
for a given $J_x$. This in turn results in an increase in $EDP_{prop}$
as seen from Fig.~\ref{fig:DSEprop}(b).

With this experiment, we obtain the design point that provides the best 
$EDP_{prop}$ as: $M_{S,\text{PMA--FM}}
= 1\times10^5$ A/m, $K_{U,\text{PMA--FM}} = 8\times10^5$ J/m$^3$, and 
$\alpha = 0.25$. At this design point, we have $v_{x} = 420$ m/s,
$t_{prop} = 480$ ps, $E_{prop} = 8.5$ aJ, $p = 1$, and $EDP_{prop} =
4.1\times10^{-27}$ Js. 

\subsection{Skyrmion nucleation}
\label{sec:skynuc}
The critical current density required for skyrmion nucleation primarily 
depends on the choice of $\alpha$~\cite{sampaio2013nucleation}. A
smaller (larger) value of $\alpha$ results in lower (higher) critical 
current density. However, as seen in
Section~\ref{sec:skyprop}, the choice of $\alpha \le 0.15$ and $\alpha =
0.2$, $K_{U,\text{PMA--FM}} = 10\times10^5 \text{J/m}^3$ are infeasible 
for skyrmion propagation. Therefore, we restrict the design space 
exploration of our nucleation process to feasible design points for 
skyrmion propagation. We perform our nucleation experiment in OOMMF
and  observe that once $J_{nuc} \ge J_{c,nuc}$, the skyrmion 
nucleates within $t_{nuc} = 20$ ps.  

In order to derive PMA--FM material parameters that provides the best
performance, we choose the design point that provides the 
minimum energy-delay product for both the nucleation and the propagation. 
This design point is obtained as: $M_{S,\text{PMA--FM}} = 1\times10^5$ A/m, 
$K_{U,\text{PMA--FM}} = 8\times10^5$ J/m$^3$, and $\alpha = 0.2$.  
At this point, for propagation we obtain $v_x = 513$ m/s, 
$t_{prop} = 389$ ps, $E_{prop} = 10.8$ aJ, and 
$EDP_{prop} = 4.2\times10^{-27}$ Js. For nucleation, the parameter
values are as follows: $I_{nuc} = 300\mu$A ($J_{nuc} = 9.5\times10^{12}$
A/m$^2$), $t_{nuc} = 20$ ps, $E_{nuc} = 3.1$ fJ, and $EDP_{nuc} = 6.2\times10^{-26}$
Js.

\subsection{Skyrmion detection and logic cascading}
We analyze the skyrmion detection and 
cascading of two SkyLogic devices using the circuit shown in 
Fig.~\ref{fig:eqDetectionCkt}(b). We perform the circuit simulation in SPICE 
with the parameters shown in Table~\ref{tbl:parameters}. 
When $T_{read}$ is turned on by applying $V_{read}= -1$V, the voltage 
at the output node is given by $V_{out} = 0.55$V
($V_{out} = 0.44$V) when $R_{\text{MTJ,0}} = 4k\Omega$
($R_{\text{MTJ,1}} = 2.5k\Omega$). Correspondingly, the next stage transistor
on current is given by $I_{on} = 300\mu$A ($I_{on} = 184\mu$A). We 
observe that $I_{on} = 300\mu$A, corresponding to $J_{nuc} =
9.5\times10^{11}$ A/m$^2$,  can nucleate a skyrmion in the next stage.
This can be seen from the OOMMF simulation results shown in Section~\ref{sec:skynuc}. 
In the case of $I_{on} = 184\mu$A, the charge current is insufficient 
for skyrmion nucleation. The circuit simulation is run for a period of 
$t_{det}=25$ ps, the time required to nucleate a skyrmion in the next stage. 
From the simulation, we determine, $E_{det}= 3.9$ fJ and $EDP_{det} =
9.7\times 10^{-26}$ Js. 

\subsection{Total delay and energy of the SkyLogic device}
\label{sec:totalDelay}
With the parameters shown in Table~\ref{tbl:parameters}, we obtain $E_{TX} =
0.1$ fJ. Using Equation~\eqref{eq:performance}, we obtain
$T_{\text{SkyLogic}} = 434$ ps and
$E_{\text{SkyLogic}} = 7.1$ fJ. The total energy-delay 
product of the SkyLogic device is given by $EDP_{\text{SkyLogic}} =
E_{\text{SkyLogic}}T_{\text{SkyLogic}} = 3.1 \times 10^{-24}$ Js.
We note that for an optimized SkyLogic inverter, as seen from
Section~\ref{sec:skynuc}, skyrmion nucleation is an energy-expensive 
process. This is due to the high current density required to nucleate
the skyrmion. The detection process, similarly also requires a high current     
through the MTJ read stack, and therefore consumes a large amount of
energy. 

\begin{figure}[ht]
\subfigure[] {
	\includegraphics[width=2.8cm]{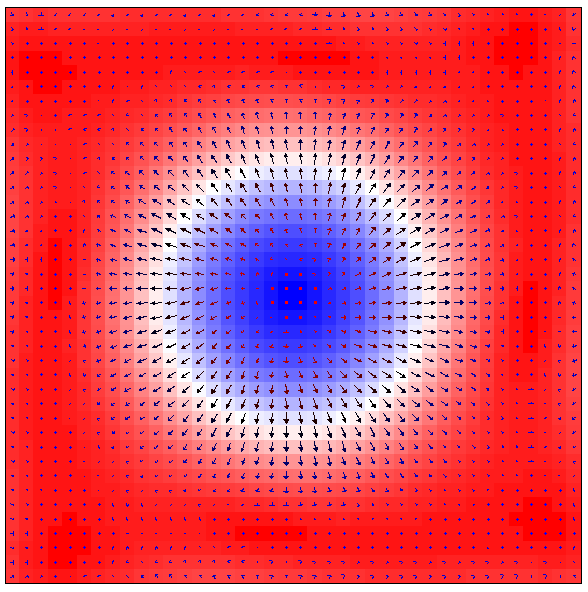}
}
\subfigure[] {
	\includegraphics[width=6cm]{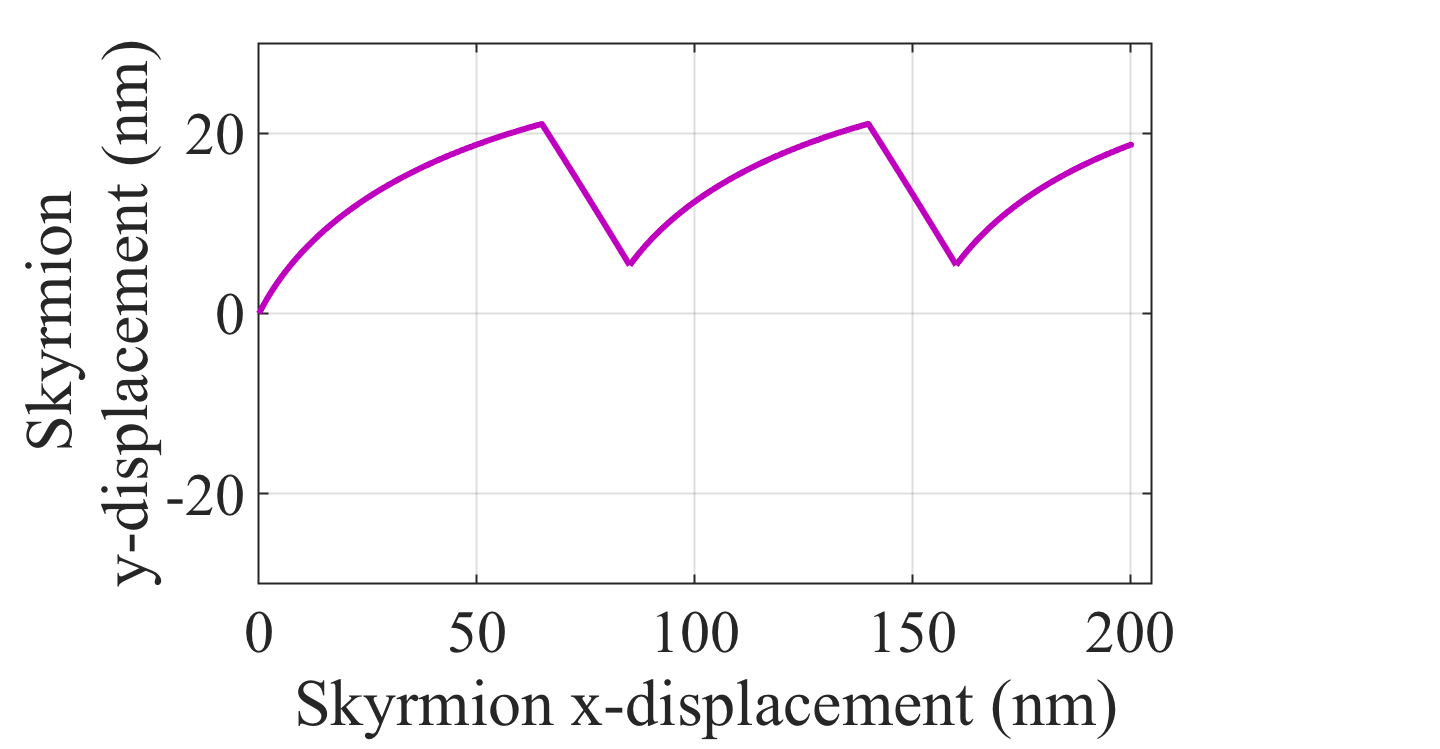}
}
\caption{At the optimal design point (a) skyrmion nucleation 
in the PMA--FM simulated in OOMMF and (b) skyrmion trajectory in the
PMA--FM. For OOMMF simulation, a spin-polarized charge current is sent
through a $20$nm diameter within a $50$nm$\times50$nm$\times0.4$nm
PMA--FM.}
\label{fig:optpoint}
\end{figure}

We show the result of the skyrmion 
nucleation for the optimal design point in Fig.~\ref{fig:optpoint}(a). 
The skyrmion trajectory in the PMA--FM during the skyrmion propagation
process for this design point is
shown in Fig.~\ref{fig:optpoint}(b). We assume that the skyrmion nucleation
process nucleates the skyrmion in the PMA--FM at (x,y) coordinate
of the skyrmion center corresponding to ($0$nm, $0$nm). It is then 
propagated to the output end of the PMA--FM such that its final (x,y) 
coordinate corresponds to ($200$nm, $18$nm). Along with the longitudinal
motion, the transverse motion of the skyrmion in $R1$ and its subsequent
motion into the interior of the PMA--FM in $R2$ can be clearly deduced
from Fig.~\ref{fig:optpoint}(b). At the optimal design point, we need two
R--SHMs to be inserted to avoid the skyrmion being destroyed at the 
edge of the PMA--FM. 

\section{Conclusion}
In this paper, we present SkyLogic, a proposal for a skyrmion-based logic device. 
We present a framework for analyzing the device performance by 
analyzing the skyrmion nucleation in a PMA--FM using micromagnetic simulation, analyzing the 
skyrmion propagation in the PMA--FM using an analytical model, and analyzing the skyrmion 
detection at the output using an MTJ stack. The transverse motion of skyrmion due to Magnus force 
during the current-driven skyrmion propagation restricts the use of high current densities 
and leads to high propagation latencies. We present a novel circuit-based technique  
to counter the Magnus force and allows the use of high current densities to 
drive the skyrmion from the SkyLogic input to the output. 
We also perform a complete design space exploration of the device for nucleation and propagation
over a range of PMA--FM material parameters and obtain an optimal design point for the SkyLogic inverter. 
At this design point, we evaluate the performance of the SkyLogic inverter and obtain a delay of $434$ ps and 
an energy of $7.1$ fJ.

\IEEEpeerreviewmaketitle

\bibliographystyle{IEEEtran}
\bibliography{ref.bib}

\begin{IEEEbiography}[{\includegraphics[width=1in,height=1.3in,clip,keepaspectratio]{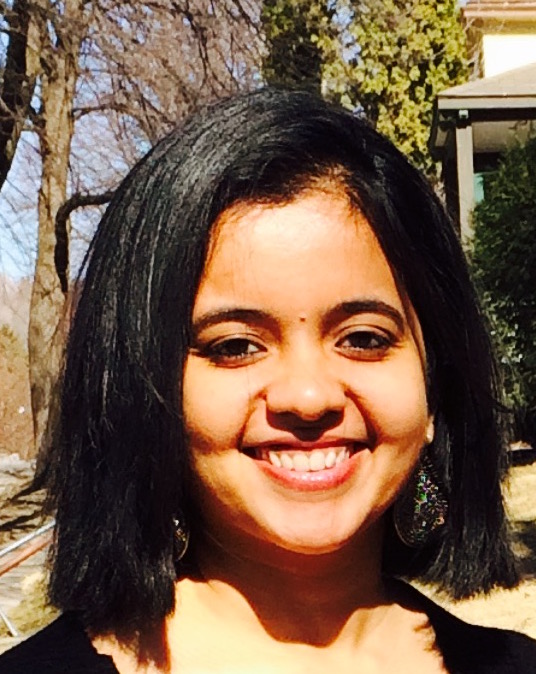}}]{Meghna
G. Mankalale} received the B.E. degree from
Visvesvaraya Technological University, India in 2007. She worked as a
Research and Development Engineer in the Electronic Design Automation
group in IBM, India from 2007 to 2013. She is currently pursuing the
Ph.D. degree in the Department of Electrical and Computer Engineering at the
University of Minnesota. Her research interests include design and
optimization techniques for beyond-CMOS technologies.  
\end{IEEEbiography}

\begin{IEEEbiography}[{\includegraphics[width=1in,height=1.3in,clip,keepaspectratio]{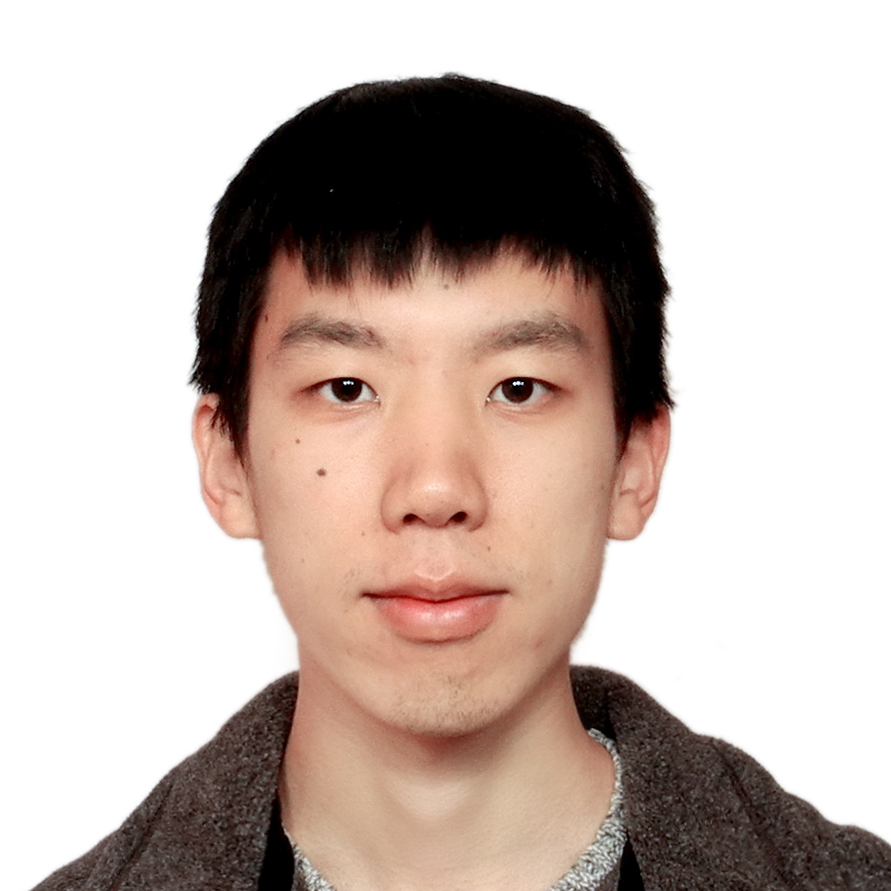}}]{Zhengyang
Zhao}
is currently pursuing the Ph.D. degree in Electrical and Computer
Engineering at the University of Minnesota, Minneapolis, MN. He received the B.S.
degree in Electrical Engineering from Xi’an Jiaotong University, China.  His
research focuses on the development of novel spintronic devices to implement
energy-efficient memory cells and logic applications. His recent work
includes studying current-induced magnet reversal using spin-orbit torque (SOT),
as well as voltage-induced magnet reversal using piezoelectric strain or VCMA
effect. More specific work includes the stack design, MTJ cell nanofabrication,
advanced device characterization and physics study.
\end{IEEEbiography}

\vskip -2.1\baselineskip plus -1fil

\begin{IEEEbiography}[{\includegraphics[width=1in,height=1.3in,clip,keepaspectratio]{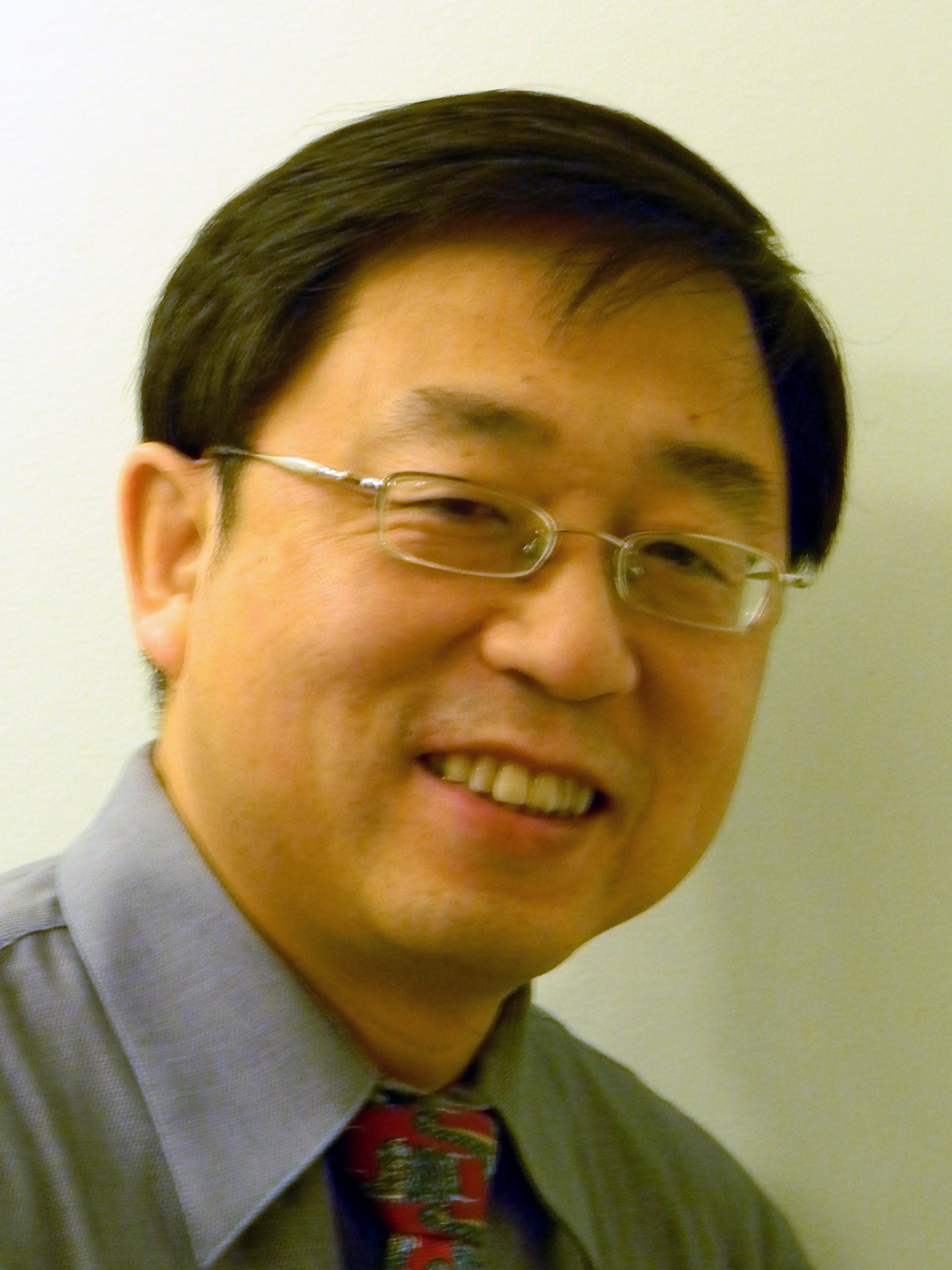}}]{Jian-Ping
Wang} received the Ph.D. degree from the Institute of Physics, Chinese Academy
of Sciences, where he performed research on nanomagnetism, Beijing, China,
in 1995. He was a Post-Doctoral Researcher with the National University of
Singapore, Singapore, from 1995 to 1996. He established and managed the
Magnetic Media and Materials Program with the Data Storage Institute,
Singapore, from 1998 to 2002. He joined the faculty of the Electrical
and Computer Engineering Department with the University of Minnesota,
Minneapolis, MN, USA, in 2002 and was promoted to Full Professor in 2009. He is the
Robert F. Hartmann Chair and a Distinguished McKnight University Professor of
Electrical and Computer Engineering and a member of the Graduate Faculty
in Physics and Chemical Engineering and Materials Science at the University
of Minnesota. He is the Director of the Center for Spintronic Materials,
Interfaces and Novel Architectures (C-SPIN), one of six STARnet program
centers. Dr. Wang received the Information Storage Industry Consortium
Technical Award in 2006 for his pioneering experimental work in exchange
coupled composite magnetic media and the Outstanding Professor Award for
his contribution to undergraduate teaching in 2010. His group is also known
for several important experimental demonstrations and conceptual proposals,
including the perpendicular spin transfer torque device, the magnetic
tunnel junction-based logic device and random number generator, ultrafast
switching of thermally stable MTJs, topological insulator spin pumping at room
temperature, and a computation architecture in random access memory.
\end{IEEEbiography}

\vskip -2.1\baselineskip plus -1fil

\begin{IEEEbiography}[{\includegraphics[width=1in,height=1.3in,clip,keepaspectratio]{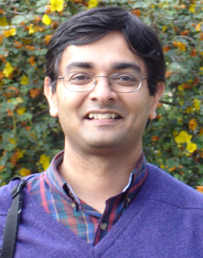}}]{Sachin
S. Sapatnekar}
(S'86, M'93, F'03) received the B. Tech. degree from the Indian
Institute of Technology, Bombay, the M.S. degree from Syracuse University, and the
Ph.D. degree from the University of Illinois.  He taught at Iowa State
University from 1992 to 1997 and has been at the University of Minnesota since
1997, where he holds the Distinguished McKnight University Professorship and the
Robert and Marjorie Henle Chair in the Department of Electrical and Computer
Engineering. He has received seven conference Best Paper awards, a Best Poster Award,
two ICCAD 10-year Retrospective Most Influential Paper Awards, the SRC
Technical Excellence award and the SIA University Researcher Award. He is a Fellow
of the ACM and the IEEE.
\end{IEEEbiography}
\end{document}